\documentclass[aps,prl,twocolumn,groupedaddress,showpacs]{revtex4}
\usepackage{graphicx}% Include figure files
\usepackage{dcolumn}% Align table columns on decimal point
\usepackage{bm}% bold math

\begin{document}

%\preprint{}

\title{Coherent Single Charge Transport in Molecular-Scale Silicon Nanowire Transistors}


\author{Zhaohui Zhong$^\dag$, Ying Fang$^\dag$, Wei Lu, and Charles M. Lieber}

\email{cml@cmliris.harvard.edu}
\altaffiliation{$^\dag$These authors contributed equally to this work.}
%\thanks{}
\affiliation{Harvard University, Cambridge, Massachusetts 02138}


\date{\today}

\begin{abstract}
We report low-temperature electrical transport studies of molecule-scale silicon nanowires. Individual nanowires exhibit well-defined Coulomb blockade oscillations characteristic of charge addition to a single structure with length scales of at least 400 nm. Further studies demonstrate coherent charge transport through discrete single particle levels extending the whole devices, and show that the ground state spin follows the Lieb-Mattis theorem. In addition, depletion of the nanowires suggests that phase coherent single-dot characteristics are accessible in a regime where correlations are strong.
\end{abstract}


\maketitle

Studies of carbon nanotubes~\cite{Yao:2001} and semiconductor nanowires~\cite{Morales:1998} have demonstrated their potential as building blocks for nanoscale electronics. An advantage of these building blocks versus nanostructures fabricated by 'top-down' processing is that, critical nanoscale features are defined during synthesis, which can yield uniform structures at the atomic scale.  Indeed, isolated carbon nanotube transistors have shown exceptional properties~\cite{Javey:2003}, although difficulties in preparing pure semiconductor nanotubes make large scale integration challenging. Silicon nanowires (SiNWs) could overcome issues faced by nanotubes since current growth methods enable reproducible control over both size and electronic properties of the nanowires~\cite{Cui:2001a,Cui:2001b}. Recent studies have begun to elucidate fundamental transport properties of chemically-synthesized semiconducting nanowires~\cite{Franceschi:2003,Thelander:2003}, which are required to move beyond initial room-temperature devices assembled with SiNWs~\cite{Cui:2001b,Huang:2001};  however, similar fundamental studies of SiNWs have not been reported. In this Letter we address this critical issue through low-temperature electrical transport measurements on molecular-scale SiNWs configured as single-electron transistors (SETs).

Single crystal p-type SiNWs with crystalline core diameters of 3 to 6~nm were synthesized by gold-nanocluster mediated vapor-liquid-solid growth using silane and diborane, and devices based-on individual SiNWs were fabricated on oxidized silicon substrates using electron beam lithography~\cite{Cui:2001a,note1}. Current ($I$) versus gate voltage ($V_g$) data recorded with a 0.5~mV source-drain bias ($V_{sd}$) at 4.2~K from a device with  source-drain separation of 400~nm exhibit regular oscillations in $I$ over a broad range of $V_g$ as shown in Fig.~1(a). The current peaks are separated by regions of zero conductance with an average peak-to-peak separation of $0.015 \pm 0.001$~V. The heights of the observed peaks vary with $V_g$, although this variation has no obvious periodicity, whereas the position and heights of the peaks are very reproducible on repeated $V_g$ scans in this and similar devices. These observations indicate that the results are intrinsic to transport through the SiNWs, and moreover, are consistent with Coulomb blockade (CB) phenomena resulting from single charge tunneling through a single quantum structure (e.g., the SiNW) with discrete energy levels~\cite{Grabert,Kouwenhoven}.  To define better the length-scale of the SiNW structure responsible for the CB oscillations, differential conductance ($\partial I/\partial V_{sd}$) versus $V_{sd}$ and $V_g$ was measured for the device in Fig.~1(a). These data (Fig.~1(b)) exhibit 33 CB diamonds, where transport is ''blocked'' for values of $V_{sd}$ -$V_g$ in the light-colored regions. The regular closed diamond structure provides strong evidence for transport through a single quantum structure and not multiple quantum dots connected in series, which would exhibit a more complex overlapping diamond structure~\cite{Kouwenhoven}. Analysis of these results~\cite{note2} yields values for the gate capacitance, $C_g$, and gate coupling factor, $\alpha  = C_g/C$, where $C$ is the total capacitance, of 10.7~aF and 0.33, respectively.

\begin{figure}
\includegraphics[width=8.2cm]{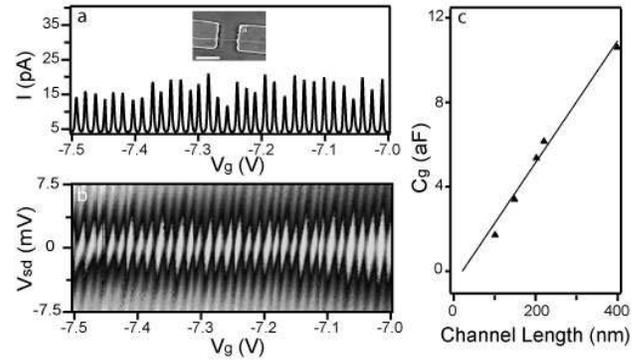}%
\caption{(a) Coulomb blockade oscillations observed at 4.2~K with $V_{sd} = 0.5$~mV. A total of 33 peaks were observed within the 0.5~V range of $V_g$. Inset: scanning electron microscopy (SEM) image of the device. Scale bar is 500~nm. (b) Grayscale plot of $\partial I/\partial V_{sd}$ vs. $V_{sd}$ and $V_g$ recorded at 4.2~K; the light (dark) regions correspond to low (high) values of $\partial I/\partial V_{sd}$; the dark color corresponds to 3000~nS.  (c) Gate capacitance vs. source-drain separation (channel length) for 5 representative devices showing single-island Coulomb blockade behavior. The line corresponds to a fit to the data with a slope of $28 \pm 2$~aF/$\mu$m.}
\end{figure}

Data exhibiting closed diamonds consistent with transport through single QDs were obtained on small-diameter SiNW devices with source-drain separations ranging from ca. 100 to 400 nm. Importantly, these data show that $C_g$ scales linearly with source-drain separation (Fig.~1(c)),   and moreover, the average value of $C_g$ determined from the data, $28 \pm 2$~aF/$\mu$m, agrees well with that calculated for a cylinder on plane model~\cite{Martel:1998}. These results supports our suggestion that the relevant dot size is defined by source-drain electrodes, since a QD size-scale set by structural variations or dopant fluctuations would give a smaller capacitance value, and be independent of the source-drain separation. The gate capacitance does deviate from the estimated value when the channel length is $<100$~nm, due to screening from the source/drain electrodes when the channel length becomes comparable to the thickness of the gate dielectric.

In addition, the variations in the current peak height versus $V_g$ in Fig.~1(a) suggest the formation of coherent energy states in the SiNW devices with energy level spacing,  $\Delta E$, larger than the thermal energy $k_BT$, where peak heights are determined by the coupling of the individual quantum states to the metal contacts at the Fermi level~\cite{Grabert}. To investigate this point further we characterized $\partial I/\partial V_{sd}$-$V_{sd}$-$V_g$ at higher resolution for a 3~nm diameter SiNW device with a 100~nm source-drain separation as shown in Fig.~2(a). The data exhibit well-defined peaks in $\partial I/\partial V_{sd}$ that appear as lines running parallel to the edges of the CB diamonds, and consistent with discrete single particle quantum levels extending across the SiNW. Analysis of the data yields $\Delta E$ values for the first 6 levels of 2.5, 1.9, 3.0, 2.0, 2.0, and 2.9~meV. These can be compared to  $\Delta E$ estimated using a one-dimensional (1D) hard wall potential: $\Delta E = (N/2)\hbar^2 \pi^2/m^*L^2$, where $N$ is the number of holes, $m^*$ is the silicon effective hole mass , and $L$ is the device length. $\Delta E$ estimated with this model ($N \approx 25$~\cite{note3}, $m^* = 0.39m_e$~\cite{Green:1990}, $L = 100$~nm), 2.5~meV, agrees well with the observed values. We also note that the level spacing is ca. 100x smaller than the sub-band spacing for these small diameter SiNWs~\cite{note4}. Similar results have been observed in more than 10 SiNW devices with source-drain separations up to 200~nm at 1.5~K, and thus we believe it is a robust feature of the small diameter SiNW devices.

\begin{figure}
\includegraphics[width=8.2cm]{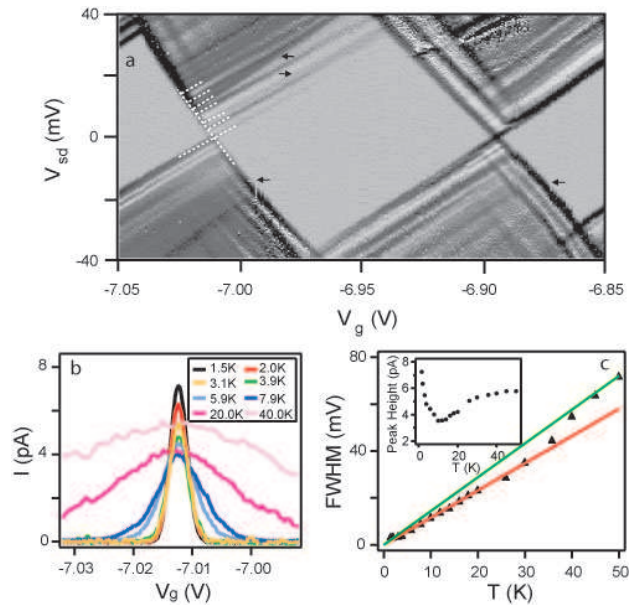}%
\caption{(a) $\partial I/\partial V_{sd}$-$V_{sd}$-$V_g$  data recorded at 1.5~K. Dark lines (peaks in $\partial I/\partial V_{sd}$) running parallel to the edges of the diamonds correspond to individual excited states, and are highlighted by white-dashed lines. Line slope changes are indicated by black arrows. (b) Temperature-dependant $I$-$V_g$ curves recorded with $V_{sd} = 50 \mu V$ at increasing temperatures. (c) Conductance peak widths in (b) determined from the full-width at half-maximum of the peak height, vs. temperature. Solid lines correspond to theoretical predictions for peak widths vs. temperature in quantum regime ($\Delta E > k_BT$),  $\alpha W=3.52ek_BT$ (red), and classical regime ($\Delta E < k_BT$), $\alpha W=4.35ek_BT$ (green), with  $\alpha = 0.26$. Inset: temperature dependence of the conductance peak height.}
\end{figure}

Temperature dependent $I$-$V_g$ measurements of the conductance peaks were also carried out. The representative data in Fig.~2(b) shows that peak current decreases rapidly as the temperature is increased from 1.5 to 10~K and is approximately constant above 30~K, consistent with coherent tunneling through a discrete SiNW quantum level that is resonant with the Fermi level of the metal contacts~\cite{Grabert,Kouwenhoven}. Moreover, the temperature at which the peak becomes constant, 30~K, yields an estimate of $\Delta E \approx 3$~meV that agrees with the value determined from the data and 1D model (see above).  In addition, the temperature dependence of the conductance peak width ($W$) is related to the gate coupling factor, $\alpha$ , as  $\alpha W=3.52k_BT/e$ in the quantum regime, $k_BT < \Delta E$, and as  $\alpha W=4.35k_BT/e$ in the classical regime, $\Delta E <k_BT< U$~\cite{Grabert}. Notably, the value of $\alpha$  determined from the temperature dependent data, 0.26, is consistent with that (0.26) obtained directly from Fig.~2(a), and thus further supports our interpretation of these experiments.

Coherent transport through a single island over length scales of several hundred nanometers indicates that these synthesized SiNWs are clean systems with little/no structural/dopant variation. Indeed, high-resolution transmission electron microscopy shows that the SiNWs have a roughness of only ca. 1-2 atomic layers on 100~nm scale, which is much less than that produced during the processing used to define the widths of planar NWs. We also speculate that dopant introduced during nanowire growth may be driven to the surface of these molecular scale SiNWs as reported for semiconductor nanocrystals~\cite{Mikulec:2000}. In contrast, low-temperature studies of nanowires with widths as small as ca. 10~nm fabricated by lithography on doped silicon-on-insulator substrates have been interpreted in terms of serially-connected quantum dots arising from variations in the potential due to structural and/or dopant fluctuations that are intrinsic to these fabricated structures~\cite{Tilke:2001}. The length scale of the electronically-distinct regions in these fabricated nanowires is on order of 10~nm, and thus ca. an order of magnitude smaller than our 3-6~nm diameter SiNWs. 

\begin{figure}
\includegraphics[width=8.2cm]{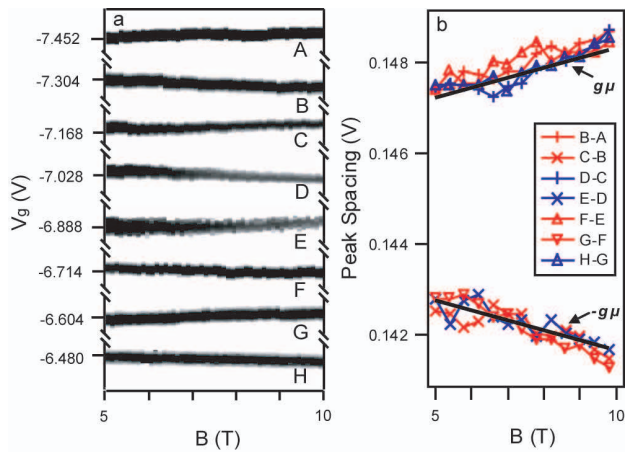}%
\caption{(a) Grayscale plot of the I as a function of $V_g$ and parallel magnetic field $B$ at 1.5~K. Data taken from the same device in Fig.~2 and 3 at a different cooling cycle. (b) Peak spacings from (a), offset to align into two branches. Solid black lines indicate expected $B$-dependence for spin transition from $\frac{1}{2}$ to $-\frac{1}{2}$ (upward), and from $-\frac{1}{2}$ to $\frac{1}{2}$ (downward). }
\end{figure}

Clean 1D systems offer unique platforms to study interactions in low-dimension systems. For example, the ground-state (GS) of QDs defined in carbon nanotubes~\cite{Cobden:1998} was found to have the lowest possible spin, while recent studies on QDs defined in two-dimensional electron gas (2DEG) showed that higher-spin ground states might also be possible~\cite{Folk:2001}. The GS spin states of the SiNW devices were studied with magnetic field parallel to the nanowire axis to minimize orbital effects. Fig.~3(a) shows the gray scale plot of $I$ as a function of $V_g$ and magnetic field $B$ taken from the same device as in Fig.~2. A small bias voltage (0.1~mV) was used so that only the ground states contribute to transport. According to Lieb and Mattis~\cite{Lieb:1962}, the GS spin in strictly 1D systems should alternate between $S = 0$ and $S = \frac{1}{2}$. As a result, the addition energy as measured by CB peak position will exhibit opposite slopes for adjacent peaks as governed by the Zeeman term, $-g\mu_B B\Delta S_Z$, with $\Delta S_Z$ alternating between $\frac{1}{2}$ and $-\frac{1}{2}$ for adjacent charge states~\cite{Cobden:1998,Folk:2001}. Indeed, data taken from 8 consecutive charge states appear as 4 down-up pairs, as shown in Fig.~3(a). The slope of peak positions as a function of magnetic field is in consistent with the Zeeman term, giving an average $g$ value of $2.0 \pm 0.2$, which agrees with the bulk Si value~\cite{Feher:1960}.  Furthermore, peak spacings extracted from the data in Fig.~3(a) are clearly divided into two branches (Fig.~3(b)): An upward branch with slope of $g\mu$ corresponding to transition from spin $\frac{1}{2}$ to $-\frac{1}{2}$ states, and a downward branch with slope of $-g\mu$ corresponding to transition from spin $-\frac{1}{2}$ to $\frac{1}{2}$ states. We did not observe a middle branch, which would be indicative of higher ground-state spin configurations, similar to observation of carbon nanotubes~\cite{Cobden:1998}. Furthermore, the simple GS spin configuration also suggests degeneracy between heavy and light holes is lifted, due to both strain and confinement effects. 

The SiNW transport data also exhibits features not explained by the constant interaction (CI) model, which we highlight in Fig.~2(a) with arrows. Specifically, we find transition lines representing ground and excited states can show different slopes, indicating that their gate coupling factors $\alpha$ are different. Bends and kinks within single transition lines show both positive and negative curvature, suggesting that $\alpha$ is not constant for a single level. Similar behavior has been observed in carbon nanotube QDs, and was attributed to either resonant defects or many-body effect~\cite{Cobden:2002, Tans:1998}.  The positive and negative slope changes seen in our data contrast expectations for the model involving resonant defects~\cite{Cobden:2002}, and may indicate the presence of electron-electron correlation (where $C_g$ is dependent on the device geometry and the many-body states).

\begin{figure}
\includegraphics[width=8.2cm]{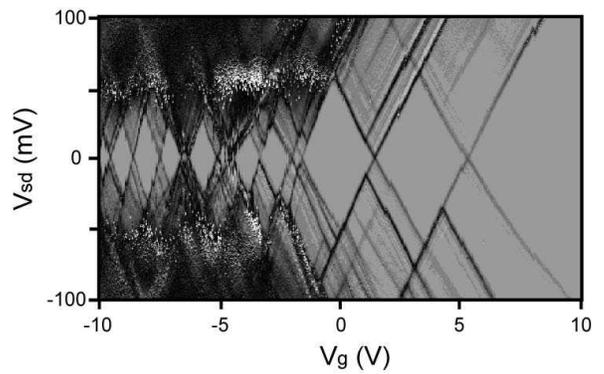}%
\caption{$\partial I/\partial V_{sd}$-$V_{sd}$-$V_g$ in grayscale for a SiNW device with 50~nm channel length at 4.2~K. The carriers are completely depleted for $V_g > 5.5$~V.}
\end{figure}

To investigate further these interesting observations we have characterized the transport properties of SiNW QDs as they are fully depleted. Representative $\partial I/\partial V_{sd}$-$V_{sd}$-$V_g$ data recorded on a 50~nm long 3~nm diameter device are shown in Fig.~4. Near $V_g = 5.5$~V, the first carrier was added to the dot. Transport is absent at more positive gate voltages, demonstrating that SiNW was fully depleted. There are several interesting features exhibited by the data in this few charge regime. First, the closed diamonds show that transport is through a single quantum structure, although the variation of diamond size, which is a measure of the minimum energy to add or remove a charge, shows substantially larger variation than data recorded in Fig.~1(b) where there are ca. 800 carriers on the SiNW QD. This variation in charging energy suggests that CI model is inadequate to treat few charge regime and that correlation may lead to shell-filling as observed previously in other QD systems~\cite{Kouwenhoven:1997}. Second, these data also show coherent tunneling through discrete SiNW quantum levels with typical level spacing of several meV. The transition lines exhibit slope changes that are more pronounced than discussed above in Fig.~2(a), and provide further evidence for charge-charge correlations. While a quantitative understanding of these observations is not yet in hand, they are nevertheless important in demonstrating the possibility of coherent transport to the few charge regime where correlations become significant. 

In conclusion, we have demonstrated that molecular-scale SiNWs can exhibit resonant tunneling at low-temperatures through discrete coherent quantum levels over length scales up to at least 200~nm. These low-temperature results exceed expectations based on many previous studies of lithographically-patterned nanowires in planar silicon, and thus point to substantial advantages of silicon-based nanowires prepared by direct synthesis versus top-down approaches for fundamental studies of 1D systems, and should serve as an interesting comparison both to carbon nanotubes and quantum wires defined in ultraclean III-V systems. Along these lines, our initial investigations of transport through molecule-scale SiNWs in the few-charge regime demonstrate rich behavior beyond the constant interaction model and could serve as a good test bed for investigating correlation effects and might ultimately have potential as a building block for quantum electronics.

We thank H.~Park, C.~Marcus, W.~Liang, S.~Datta and S.~Hareland for helpful discussion. C.M.L. is grateful for support of this work by the Defense Advanced Research Projects Agency, Intel, ARO, and NSF.

%\bibliography{Zhong}

\begin{thebibliography}{34}
\bibitem[{not({\natexlab{a}})}]{note1}
\bibinfo{note}{Single crystal silicon nanowires were prepared with
  SiH$_4$:B$_2$H$_6$ ratio of 8000:1 using 5~nm diameter gold nanocluster
  catalysts. Devices were fabricated on degenerately doped silicon substrates
  with 50~nm thermal oxide using electron-beam lithography and subsequent
  evaporation of 50~nm Ni to define source-drain contacts; devices were
  annealed at 350~$^\circ$C for 20~seconds.}
\bibitem[{not({\natexlab{b}})}]{note2}
\bibinfo{note}{The energy to add a charge onto the SiNW, $U+\Delta E$, is
  directly measured from the maximum in $V_{sd}$ for the conductance gap of the
  diamond, where $U = e^2/C$ is the charging energy; the addition energy
  determined from Fig.~1(b) is around 5~meV. The peak spacing $\Delta V_g$ is
  related to the addition energy by $\Delta V_g = (U+\Delta E)/e\alpha$ .
  Assuming that $\Delta E$ is small compared to $U$ yields values of $C_g =
  e/\Delta V_g \approx 10.7$~aF and $\alpha = 0.33$.}
\bibitem[{not({\natexlab{c}})}]{note3}
\bibinfo{note}{The number of carriers were estimated from the difference in
  $V_g$ required to deplete fully the device ($\sim 4$~V) and that of the
  measurement ($\sim -7$~V) using $N = C_g \Delta V_g/e$. For specific
  parameters in Fig.~2(a), $N \approx 25$.}
\bibitem[{not({\natexlab{d}})}]{note4}
\bibinfo{note}{The subband energy level spacing, $E_n$, was estimated using a
  model with a hard wall potential, $E_n=\frac{\hbar^2\pi^2n^2}{2m^*w^2}$,
  where $n$ is the subband number, $m^*$ is the effective mass, and $w$ is the
  nanowire diameter. For 3~nm diameter nanowires and $m^* = 0.39m_e$, the
  spacing between $n = 1$ and 2 is 320~meV. The Fermi energy, $E_F$, in 1D is
  given by $E_F=(\frac{Nh}{4\sqrt{2m^*}})^2$ , where $N$ is the density of
  charges per unit length. For typical value of $N$ at 4.2~K, $25/100$~nm, $E_F
  = 15$~meV, implying that only the first subband is occupied.}
\end{thebibliography}

\end{document}